\documentclass[fleqn,usenatbib]{mnras}

\usepackage{newtxtext,newtxmath}
\usepackage[T1]{fontenc}
\DeclareRobustCommand{\VAN}[3]{#2}
\let\VANthebibliography\thebibliography
\def\thebibliography{\DeclareRobustCommand{\VAN}[3]{##3}\VANthebibliography}

\usepackage{times}%

\usepackage{graphicx}	
\usepackage{colordvi}
\usepackage{amsmath}

\usepackage{amssymb}
\usepackage[toc,page]{appendix}
\usepackage{natbib}
\usepackage{array}
\usepackage{comment}

\usepackage{lineno}
\usepackage[usenames,dvipsnames]{xcolor}

\title[Astrometric detection of binary asteroids]{Astrometric detection of binary asteroids}
\author[Segev, Ofek \& Polishook]{
Noam Segev\thanks{E-mail: noam.segev@weizmann.ac.il},
Eran O. Ofek
and David Polishook
\\
Department of Particle Physics and Astrophysics, Weizmann Institute of Science, 76100 Rehovot, Israel.
}

\begin{document}
\newcommand{\gsim}{\raisebox{-0.13cm}{~\shortstack{$>$ \\[-0.07cm]
      $\sim$}}}
\label{firstpage}
\pagerange{\pageref{firstpage}--\pageref{lastpage}}
\maketitle

\begin{abstract}
	Binary asteroids probe thermal-radiation effects on the main-belt asteroids' evolution. We discuss the possibility of detecting binary minor planet systems by the astrometric wobble of the center-of-light around the center-of-mass. This method enables the exploration of the phase-space of binary asteroids, which is difficult to explore using common detection techniques. We describe a forward model that projects the center-of-light position with respect to the center-of-mass, as it is seen by the observer. We study the performance of this method using simulated Gaia-like data. We apply the astrometric method to a subset of the Gaia DR2 Solar System catalog and find no significant evidence of binary asteroids. This is likely because the Gaia DR2 removed astrometric outliers, which in our case may be due to astrophysical signals. 
	Applying this method to binary asteroid (4337) Arecibo, for which Gaia DR3 reported a possible astrometric signal with a period of $P = 32.85\pm0.38$\,hr, reveals a possible 2.2-$\sigma$ solution with a period of 16.26\,hr (about half the reported period). We find a small, marginally significant, excess of astrometric noise in the known binary asteroid population from Pravec et al. relative to the entire asteroid population in the Gaia DR2 Solar System catalog. We also discuss some caveats like precession and asteroid rotation.
\end{abstract}


\section{Introduction}
Gravitationally bound binary asteroids in the Solar System enable us to study asteroids' properties, formation, and evolution.  
Since the discovery of Dactyl \citep{galileo_Ida_sat_dactyl}, Ida's moon, the population of the known binary asteroids has grown to hundreds of systems, including tens of triplets, and recently a quadruple system was found \citep{quadruple_asteroid_berdeu2022first}.

To date, binary asteroids were detected using three main techniques (carefully described by \citealt{binary_review_2000}): (i) direct imaging from space or ground-based telescopes with adaptive optics (AO); (ii) eclipses and periodicity in the lightcurve; and (iii) radar observations.

AO systems have revealed tens of binary asteroids. Since it provides resolution of the order of 0$\mathord{''}$.1 it is limited to the separation of $\mathord{\gsim}$150\,km for binary systems in the main belt.

The bulk of the known binary asteroids ($\mathord{\sim}$300) were detected by mutual eclipse and occultation events between the components of each binary, observed in the lightcurves of these asteroids \citep{binary_asteroid_review_margot2015asteroid}. The geometric conditions for an occultation event significantly reduce the probability of detecting binary asteroids with this method. However, this is still the most effective detection method, mainly since it is the most accessible technique. Furthermore, given enough photometric observations this technique has the potential to detect {\it all} the binary asteroids. Additionally, a few tens of binary asteroids among Near-Earth Asteroids (NEAs) were detected by radar observations from Earth.

How binary systems form remains uncertain, though it presumably involves multiple formation channels and mechanisms.

\cite{jacobson2011dynamics_two_tracks} present a model that describes the creation of NEA by rotational fission induced by the Yarkovsky-O’Keefe-Radzievskii-Paddack (YORP) effect \citep{yorp_review_bottke2006yarkovsky}. This model suggests two evolutionary tracks, distinguished by the secondary- to-primary-mass ratio. The evolution tracks differ by the post-fission {\it free energy}, the available energy for rotation and orbit in the binary system, which is defined as the kinetic energy plus the mutual potential energy of the components \citep{scheeres2006dynamics_free_energy_def}.  The free energy of two spherical components with equal density as a function of the secondary-to-primary-mass ratio is a monotonically decreasing function that equals zero for $q\approx0.2$ (diameter ratio $D_2/D_1\mathord{\approx} 0.58$; \citealt{scheeres2007rotational_fission_free_energy}). Since the free energy is negative for high-mass-ratio systems ($q>0.2$), these systems are bound. The free energy is positive for low mass ratio systems ($q<0.2$). Hence, these systems are unstable and may be disrupted or undergo secondary fission.

However, the reason for the apparent gap in the diameters ratio of $0.4<D_2/D_1<0.8$ (translates to mass ratio of $0.13<q<0.5$ for spherical components with the same density), shown in the top panel of Figure \ref{fig:Dratio_hist}, is uncertain.

In this paper, we review the possibility of detecting unresolved binary asteroids by the motion of their center-of-light around the center-of-mass (i.e., the astrometric method), and applying it to Gaia DR2 observations for which residuals from the orbital fit were published. This technique is most sensitive to binaries with a diameters ratio of $\sim$0.5, the approximate division between the two evolutionary tracks suggested by \cite{jacobson2011dynamics_two_tracks}. Although this method is heavily used in other fields of astronomy (e.g.,  \citealt{wobble_planet_neuhaeuser2006detectability}, \citealt{boss2009carnegie_wobble}, \citealt{wobble_gaia_belokurov2020unresolved}, \citealt{springer2021measuring_a,springer2021measuring_b} etc.), it is hardly discussed or used in the asteroids literature.

The Gaia Data Release 3 (DR3) Solar System Object \citep{tanga2022data_GAIA_DR3_SSO}, which was just released, demonstrates the capability of Gaia DR3 to measure the binary astrometric wobble of the main belt asteroid (4337) Arecibo, using observations that span over 2.3 days. This demonstration supports the possibility to detect new asteroids by the astrometric method, which enables the use of the full time span of Gaia observations.

In \S \ref{sec:physical_model}, we write the physical model for the center-of-light wobble, and in \S \ref{subsec:forward_model}, we derive the forward model. In \S\ref{sec:model_inversion}, we describe the model inversion algorithm, while in \S\ref{sec:data_collection}, we discuss the Gaia data. In \S\ref{sec:simulations}, we test the sensitivity of the algorithm under several scenarios, and in \S \ref{sec:results}, we apply our method to 20 selected objects from the Gaia DR2 Solar System Objects (SSO) catalog. Finally, in \S \ref{sec:caveats}, we discuss the caveats of the astrometric method, and in \S\ref{sec:conclusion}, we summarize our conclusions.

\section{Astrometric wobble of the center of light}
\label{sec:physical_model}
The center-of-mass position of a binary asteroid depends on the ratio of the masses of the components. In contrast, the center-of-light depends on the asteroid’s projected surface ratio as seen by an observer and on the surface albedo. Therefore, in the case of binary asteroid components with nonidentical diameters, the center of light will deviate from the center-of-mass position.
This deviation causes the center-of-light of a binary asteroid to wobble around the center-of-mass while the center-of-mass moves in a nearly Keplerian orbit around the Solar System barycenter.

Here, we derive the projected center-of-light position as seen by an observer relative to the binary center-of-mass. This forward model will enable us to invert the problem, search for binary asteroids, and measure their orbital parameters.

\subsection{Physical model }
\label{subsec:model}
In principle, fitting the center-of-light motion of a binary asteroid should be done simultaneously with fitting the center-of-mass motion around the Solar System Barycenter. However, given enough observations, the two problems can be approximately separated. Specifically, in the case of an unresolved binary asteroid, it is sufficient to fit the Keplerian orbit around the Solar System Barycenter's  apparent position, i.e., the center-of-light, as if it were a single asteroid.
For a sufficient number of observations, the wobble of the center-of-light will be averaged out\footnote{This is only approximately correct for an eccentric orbit.}. If the residuals are statistical, the amplitude of the residuals' contamination from the binary nature of the asteroid will decrease like $1/\sqrt{N_e}$, where $N_e$ is the number of epochs from which the orbit is derived.

To describe the orbital elements of the binary asteroid components around the center of mass, we use a reference frame defined by the ecliptic coordinates system and the J2000.0 vernal equinox.
Here, $\Omega$ is the longitude of the ascending node, $\omega$ is the argument of periapsis, $i$ is the inclination, and $e$ is the eccentricity. In addition, $M_0$ is the mean anomaly at some fiducial epoch, here chosen as J2014.0, $P$ is the binary orbital period, $M$ is the total mass of the binary pair, and $q\equiv m_2/m_1$ is the mass ratio, where $m$ is the mass of individual components, and the subscripts $2$ and $1$ stand for the secondary and primary components, respectively. Under the assumption of spherical components with similar density and diameters $D_1$ and $D_2$, the mass ratio is equal to the volume ratio and, therefore, can be written as $q= (D_2/D_1)^3$.

\begin{figure}\centering
\includegraphics[width=1\linewidth]{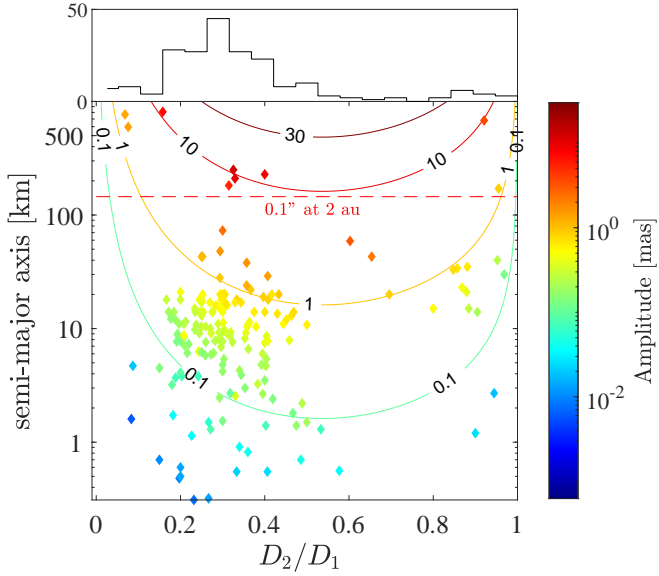}\hfill
\caption{The astrometric amplitude as a function of the semi-major axis and diameter ratio (secondary over primary) for MBA. The colored diamonds represent known MBA binaries with measured semi-major axis and diameter ratio \citep{pravec2019_asteroid_database}. The center-of-light wobble amplitude (in mas) is calculated using Equation \ref{eq:col_q}, with the Heliocentric semi-major axis used as the object-observer distance. The contours are calculated for objects at 2\,au from the observer. The red, dashed horizontal line shows the binary semi-major axis, which appears as a 0.1$''$ separation at a distance of 2\,au (objects below this separation will be unresolved by Gaia). The top histogram shows the known binaries' diameter ratio histogram, with a bin size of 0.053.} 
\label{fig:Dratio_hist}
\end{figure}

\begin{figure}\centering
\includegraphics[width=1\linewidth]{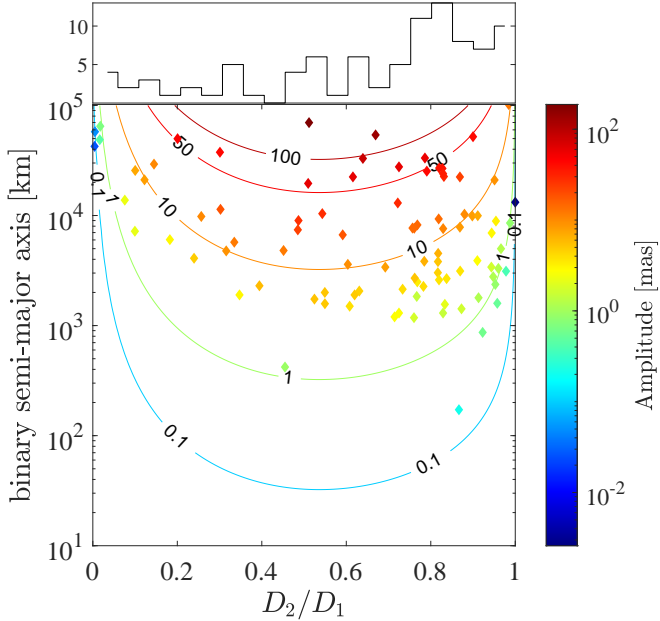}\hfill
\caption{Same as in Figure \ref{fig:Dratio_hist} but for known Trans-Neptunian Object's satellites, where the contours are calculated at 40\,au from the observer. The physical parameters were taken from \protect\cite{warner_nasa_database}.} 
\label{fig:Dratio_hist_kuiper}
\end{figure}

The apparent asteroid flux depends on the asteroid albedo and projected surface area. 
Assuming an asteroid with a uniform albedo $A$, we can write the center of light position as
\begin{align}
    \vec{x}_{col} = \frac{\vec{x}_1 A_1 D_1^2 + \vec{x}_2 A_2 D_2^2}{ A_1 D_1^2 +  A_2 D_2^2},
    \label{eq:col_full}
\end{align}
where $\vec{x}_{col}$, $\vec{x}_{2}$ and $\vec{x}_{1}$ are the center of light and the secondary and primary component positions in the binary center-of-mass frame, respectively.
Assuming the components have the same density and albedo, we can write the center-of-light position, in the center-of-mass frame, by
\begin{align}
    \vec{x}_{col} = \frac{q^{2/3}-q}{1+q+q^{2/3}+q^{5/3}} \vec{x}\equiv f(q)\vec{x},
    \label{eq:col_q}
\end{align}
where $\vec{x}= \vec{x}_2-\vec{x}_1$. The function $f(q)$, which we call the center-of-light scaling function, reaches a maximum at $q=(D_2/D_1)^{3} \cong 0.15$ (i.e., $D_2/D_1 \cong 0.53$). 
The contours in Figures \ref{fig:Dratio_hist} and \ref{fig:Dratio_hist_kuiper} show the expected angular amplitude of $x_{col}$ as a function of the semi-major axis and binary-asteroid diameter ratio for asteroids in main belt (MBAs) and Trans-Neptunian Objects (TNOs), respectively. Figure \ref{fig:Dratio_hist_kuiper} suggest that some binary TNO can be detected easily using astrometry from ground based survey (e.g., \cite{astrometry_ofek2019}).

In the case of resolved binary components, the astrometric wobble of the primary asteroid around the center of mass can be used for binary detection. The primary component position with respect to the center-of-mass is given by 
\begin{align}
 	\vec{x}_1 = -\vec{x}\frac{q}{1+q}.
\end{align}
The unresolved and resolved astrometric wobble differ only by their amplitude. Therefore, the forward model and model inversion, described in \S\ref{subsec:forward_model} and \S\ref{sec:model_inversion}, respectively, can be used in both cases to detect binary asteroids. Here, we focus on the unresolved binary case, which is more likely to be detected by Gaia.

\section{Forward model}
\label{subsec:forward_model}
In order to detect a binary asteroid and fit its orbital parameters, we write the forward model that, given the binary asteroid orbital parameters, calculates the center-of-light position as seen by an observer.

First, we calculate $\vec{x}_{col}$, the center-of-light position in the rest frame as a function of time, given a set of orbital parameters $\left(\Omega,\omega,i,M_0,e,P,q,M\right)$. 
We define the rest frame coordinate system centered on the binary asteroid center-of-mass, $\hat{\xi}$, in a right-handed heliocentric Cartesian coordinate system in which $\hat{\xi}_x$ is directed toward the vernal equinox and $\hat{\xi}_z$ is directed toward the north ecliptic pole.

Next, we calculate the observer-to-targets unit vector\footnote{After correcting for the light-travel time.} 
\begin{align}
    \hat{t} = \frac{\vec{T}-\vec{O}}{\left\lVert\vec{T}-\vec{O}\right\rVert},
\end{align}
where $\vec{T}$ and $\vec{O}$ are the target and observer vector positions relative to the Solar System barycenter, respectively, and $\left\lVert x \right\rVert$ represents the L2 norm of $x$.

Next, we rotate both $\hat{t}$ and $\vec{x}_{col}$ from the ecliptic to the equatorial coordinate system using the following rotation matrix

\begin{align}
M(\epsilon)
&= \begin{bmatrix}
    1   & 0    &0 \\
    0   & \cos \epsilon      &  -\sin \epsilon \\
    0       &\sin \epsilon           & \cos \epsilon \\
    \end{bmatrix},
\end{align}
where $\epsilon$ is the obliquity of the ecliptic. We project the center-of-light position to the plane perpendicular to the observer-asteroid (center of mass) line of sight. To do so, we project the North celestial pole component, i.e., $\hat{n}=(0;0;1)$, to this plane (i.e., perpendicular to $\hat{t}$) by 
\begin{align}
    \hat{n}_{N} = \frac{\hat{n} - (\hat{n}\cdot\hat{t})\hat{t}}{\left\lVert\hat{n} - (\hat{n}\cdot\hat{t})\hat{t}\right\rVert}.
    \label{eq:hat_nN}
\end{align}
Here, the symbol $\square\cdot\square$ represents the Scalar product. In the next step, we find the vector perpendicular to $\hat{t}$ and $\hat{n}_N$, which points toward the East by 
\begin{align}
    \hat{n}_E = \hat{n}_{N}\times \hat{t} ,
	\label{eq:hat_nW}
\end{align}
where the symbol $\square\times\square$ represents the cross product.
At this point, we calculate the center-of-light position with respect to the center-of-mass as seen from the observer-to-target line of sight. Finally, we convert the projected center of light position into angular distances in right ascension and declination by
\begin{align}
\Delta \alpha &= \arctan \left(\frac{\vec{x}_{col}\cdot \hat{n}_E}{d}\right),\\
\Delta \delta &= \arctan \left(\frac{\vec{x}_{col}\cdot \hat{n}_N}{d}\right),
\label{eq:Delta_alpha_delta}\end{align}
where $d$ is the observer-target distance. Note that $\Delta \alpha$ is measured on a great circle and in angular units (i.e., no cos$\,\delta$ term).

As described in Appendix \ref{subsec:gaia_al}, when dealing with Gaia observations, we have to apply an additional projection to the Gaia along-scan axis. Throughout the paper, we use tools from \cite{eran_matlab_ofek2014matlab}.

\section{Model inversion}
\label{sec:model_inversion}

To detect binary asteroids by the center-of-light wobble around the center-of-mass, we fit the forward model to the observed astrometric residuals from the fitted center-of-mass Solar System orbit.
We fit the wobble amplitude, in milliarcseconds, as a single parameter ($\mathcal{A}$) instead of fitting for $q$ and $M$ separately, since the amplitude is a degenerate function of those two parameters (see Equation \ref{eq:col_q}). Therefore, the orbital binary fit includes seven free parameters $\vec{\theta} \equiv(\Omega, \omega, i, M_0, P,e,\mathcal{A})$.

In the fit procedure, we enumerate over a grid of orbital periods ($P$), and for each period, we fit the other parameters by minimizing the $\chi^2$,  which is calculated by

\begin{align}
\chi^2_{axis}(\vec{\theta}) = \sum_i\frac{\left( \bar{\delta}_{i}-x_i(\vec{\theta}) \right)^2}{\sigma_i^2},
\label{eq:chi_squared}
\end{align}
where the $i$-index represents an epoch, $x(\vec{\theta})$ is the expected center-of-light wobble as seen by the observer for a set of given $\vec{\theta}$ parameters, $\bar{\delta}$ is the observed residual from the center-of-mass orbital fit around the Solar System, and $\sigma$ is the astrometric measurement uncertainty.
For the minimization, we use standard numerical solvers\footnote{Using the Nelder-Mead simplex algorithm, as described in \citealt{Nelder_mean_lagarias1998convergence}.}, i.e., {\tt MATLAB} built-in {\tt fminsearch.m}. Equation \ref{eq:chi_squared} is calculated for each axis separately, and the two $\chi^2$ are combined.

 Due to the aliasing and the complex window function, it is not trivial to estimate the significance of the fit and declare detections. Here, we present an empirical approach to estimate the significance of the fit.

To test the detection significance, we calculate the false positive rate (FPR) using the Bootstrap technique \citep{bootstrap_efron1992bootstrap}, i.e., reassigning residuals to different epochs, to estimate\footnote{In the case in which there is a real periodicity in the data, the Bootstrap will overestimate the null hypothesis $\chi^2$.} the $\chi^2(H_0)$ distribution under the null hypothesis ($H_0$) of a single asteroid model (i.e., $\mathcal{A}=0$).

To calculate the significance level of the detection, we compare the obtained $\chi^2(H_1)$ of the best fit binary model ($H_1$) with the tail of the null hypothesis distribution.
The probability density function (pdf) of a $\chi^2$ distribution with $k$ degrees-of-freedom (for $x>0$) is given by
\begin{align}
    p(x,k) = \frac{x^{\frac{k}{2}-1}e^{-\frac{x}{2}}}{2^{\frac{k}{2}} \Gamma\left(\frac{k}{2}\right)},
\end{align}
where $\Gamma$ is the Gamma function. In the limit of $x\gg k$, the $\chi^2$ pdf decays approximately as $\exp\left(-x/2\right)$, independent of the number of degrees-of-freedom.
This feature provides a robust method to calculate $\alpha$ in the case of a non-linear model, in which the number of degrees of freedom is unknown.

Therefore, to estimate the FPR, we select the $n$-percentile tail ($\chi^2_{<n}$) of the null hypothesis distribution (obtained by the bootstrap simulations) and fit an exponential distribution for each of the frequencies separately. The exponential distribution is given by:
\begin{align}
    x_0 &= \chi^2_{n} - \chi^2_{<n}\label{eq:exp_variable}\\
    x_0  &\sim \exp(\lambda), \label{eq:exp_dist}
\end{align}
where $\chi^2_n$ is the $n$-percentile of the null hypothesis, calculated for each of the frequencies separately.
For the choice of 500 bootstrap simulations, we find  $n=20$-percentile as a robust tail.

\begin{figure}\centering
	\includegraphics[width=1\linewidth]{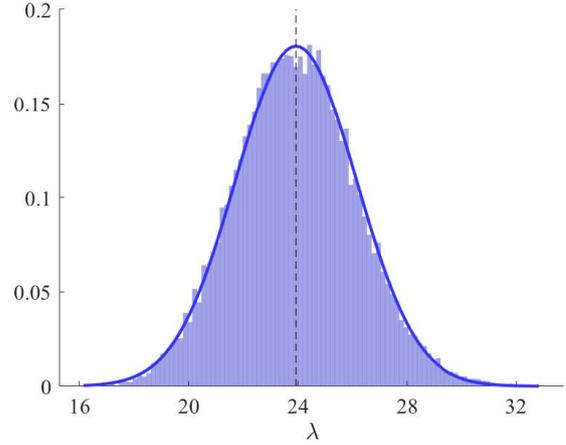}\hfill
	\caption{Histogram of the best fit of the exponential distribution parameter ($\lambda$) of the null hypothesis (Equations \ref{eq:exp_variable} and  \ref{eq:exp_dist}) for (5899) Jedicke's observations in Gaia DR2. The exponential distribution were fitted separately for each frequency in the frequency range $1/350\,\text{hr}^{-1}$ to $1/10\,\text{hr}^{-1}$ with steps of $1/505 \, \text{hr}^{-1}$. The black dashed line shows the best fit exponential parameter for $x_0$ in all of the frequencies combined ($\lambda = 23.92$), i.e., we calculate $x_0$ (Equation \ref{eq:exp_variable}) in each of the frequencies separately and fit the combined $x_0$ with a single exponential distribution.}
\label{fig:lambda_dist_jedicke}
\end{figure}

 An example of the best fit exponential parameter $\lambda$ of asteroid 5899 Jedicke observations in Gaia DR2 for the frequency range $1/350\,\text{hr}^{-1}$ to $1/10\,\text{hr}^{-1}$ with steps of $1/505 \, \text{hr}^{-1}$. In this case, we use 500 Bootstrap resampling, i.e., each frequency has 500 measurements of $\chi^2(H_0)$. The distribution of the best fit $\lambda$ of each frequency agrees with a Normal distribution, as shown in Figure \ref{fig:lambda_dist_jedicke}.

We calculate the FPR ($\alpha$) for $\chi^2(H_1)$ using the exponential cumulative distribution function, given by
\begin{align}
    x_1&= \chi^2_{20}(H_0) - \chi^2(H_1)\\
    \alpha(x_1,\lambda)& = 1 - \int_0^{x_1} {\lambda e^{-\lambda y}dy}.
\end{align}

The required significant level of $\alpha$ needs to be adjusted for each data set, since the number of independent measurements is unknown as the frequencies are correlated.

\section{The Gaia data}
\label{sec:data_collection}
In this paper, we apply the astrometric method to observations from the Gaia DR2 Solar System Observations (SSO) catalog \citep{GAIADR2_SS_2018}.
The Gaia slowly rotating and precessing spacecraft consists of two telescopes pointing about 105$^\circ$ apart, while their focal planes are projected on the same array of detectors \citep{gaia_mission_prusti2016}. In each epoch, the targets cross nine CCDs. 
In the nominal magnitude of $G\sim15$, this scanning strategy provides an astrometric precision of $\mathord{\sim} 100\,\mu$as for stellar objects per visit along the axis parallel to the target's motion on the detector, called the along-scan axis. 
The astrometric precision is roughly $\sim$0.2 arcseconds for the across-scan, i.e., the perpendicular axis. The Gaia pixel scale is 58.9×176.8\,mas\,pixel$^{-1}$ in the along-scan and across-scan directions, respectively.
Therefore, when searching for binary asteroids using Gaia DR2, we use only the along-scan astrometry.

The Gaia collaboration published the fitted orbits for 14,099 asteroids around the Solar System Barycenter in the Gaia DR2 SSO catalog, together with the residuals measured for each CCD crossing. However, Gaia did not publish the full data set, as we discuss next in this section.

\begin{figure}
	\includegraphics[width=84mm,height=60mm]{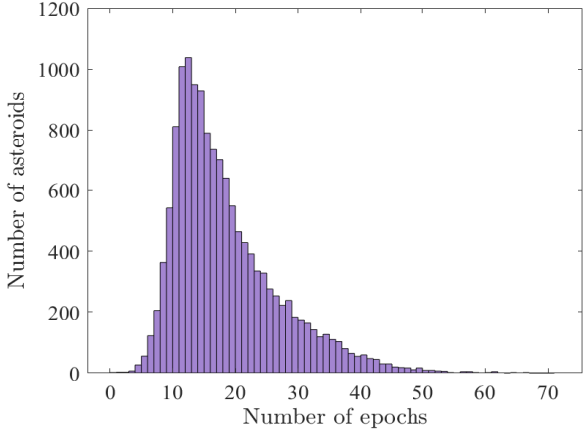}
	\caption{The number of epochs for the 14,099 asteroids in the Gaia DR2 SSO catalog, after discarding epochs with less than 4 sub-transits. We defined epoch by one transit over the focal plane, as described in \S\ref{sec:data_collection}. }
	\label{fig:epoch_num_hist}
\end{figure}

In this work, we define an epoch as one transit of the target across the telescope field of view. Each epoch contains up to nine independent CCD sub-transits. In this work, we discarded epochs in which there are less than 4 sub-transits ($\mathord{\sim}10\%$ of epochs). A histogram of the number of epochs is shown in Figure \ref{fig:epoch_num_hist}.

\begin{figure}\centering
    \includegraphics[width=0.95\linewidth,trim={0 0 0 0.5mm},clip]{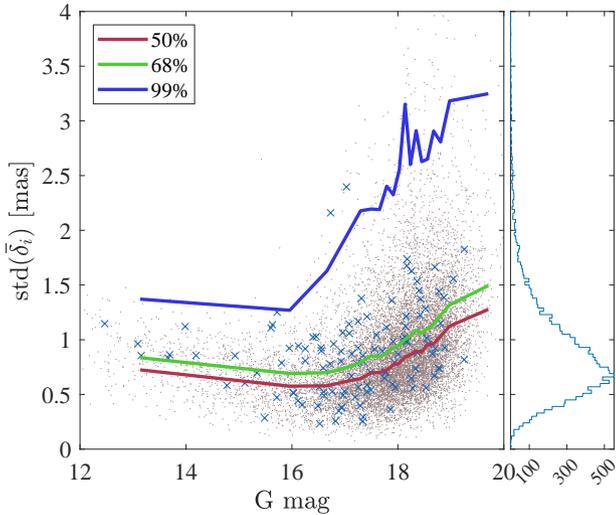}\hfill
	\caption{The Gaia DR2 SSO catalog's object standard deviation of $\bar{\delta}_i$ (i.e., scatter of sub-visit measurements in a single epoch) as a function of the apparent $G$ mag. The colored lines present the $G$-mag-binned percentiles of std($\bar{\delta}_i$): red for the median (50\%), green for 68\%, and blue for 99\%. The known binaries from \citealt{pravec2019_asteroid_database} are indicated by blue crosses.}
	\label{fig:magG_std}
\end{figure}

Let $\delta_{i,s}$ be the along-scan residual in sub-transit $s$ of epoch $i$, and $\bar{\delta}_i$ the mean residual, calculated over all sub-transits in epoch $i$. 
Figure\,\,\ref{fig:magG_std} shows the $G$ mag as a function of the standard deviation of $\bar{\delta}_i$ over all epochs, where each dot represents an asteroid from the Gaia DR2 SSO catalog and each blue cross is a known binary from the \cite{pravec2019_asteroid_database} catalog that has Gaia observations.

\begin{figure}\centering
	\includegraphics[width=0.95\linewidth,trim={0 0 0 0.5mm},clip]{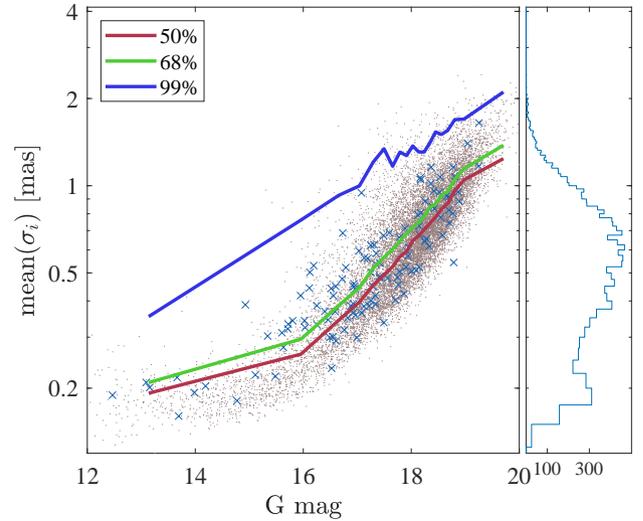}\hfill
    \caption{The objects' mean standard error ($\sigma_i$) as a function of the apparent $G$ mag from Gaia DR2 SSO catalog. The markers are like those shown in Figure \ref{fig:magG_std}.}
	\label{fig:G_prec}
\end{figure}

We calculate the standard error of $\bar{\delta}_i$ by 
\begin{align}
\sigma_i = \frac{\sqrt{\sum_s \left(\delta_{i,s} - \bar{\delta}_{i}\right)^2}}{N_i-1},
\label{eq:sigma_i}
\end{align}
where $N_i$ is the number of sub-transits in epoch $i$. Figure \ref{fig:G_prec} shows the Gaia DR2 SSO asteroids' mean standard error of $\bar{\delta}_i$ (i.e., mean of $\sigma_i$ over $i$, for each asteroid), calculated over all epochs for each asteroid, as a function of the $G$ magnitude. The asteroid standard error in the Gaia DR2 SSO catalog reaches a noise floor of $\mathord{\sim}0.4$\,mas for asteroids brighter\footnote{For objects brighter than $G$=13, Gaia readouts the entire stamp around the source. However, in order to reduce data volume, Gaia readouts only a window of 12X12 (18X12) pixels around each object that is fainter than $G=16$ (G=13), and bins the pixels perpendicular to the scanning direction (see \citealt{GAIADR2_SS_2018} )} than $G$ magnitude 16.
For fainter asteroids, with $G>16$, the standard error increases as the brightness decreases, as is expected for Poisson noise. A summary of Gaia DR2 SSO observations with a detailed description of the astrometric uncertainties and systematics can be found in \cite{GAIADR2_SS_2018}.

Unfortunately, the published Gaia DR2 SSO catalog observations do not contain outliers that were clipped during the orbital fit procedure. One of the rejection criteria is transits in which the mean residuals ($\bar{\delta}_i$ in this work) of the Solar System barycenter orbital fit is higher than the systematic error (presumably $\sigma_i$ in this work, but we did not find an exact definition in \citealt{GAIADR2_SS_2018}). These criteria rejected observations with large residuals, even though they are consistent between the epoch sub-transits (see {\it Validation of the astrometry} section in \citealt{GAIADR2_SS_2018}).
These rejected transits do not appear in the Gaia DR2 SSO catalog. Unfortunately, this fact severely limits our ability to use the Gaia DR2 catalog for binary detection. Nevertheless, we attempted to search for such binaries in \S\ref{sec:results}.

We check whether the known binaries' sample \citep{pravec2019_asteroid_database} shows an excess signal in the residuals from the Solar System barycenter orbital fit, compared to the rest of the Gaia DR2 SSO catalog (Gaia sample).
To do so, we use the two-sample Kolmogorov--Smirnov test (KS-test, \citealt{kolmogorov_test_massey1951}), where the null hypothesis is that the $\text{std}(\bar{\delta}_i)$ of both the known binaries and the Gaia sample are from the same continuous distribution. The alternative hypothesis is that the samples are from different continuous distributions. We bin the two samples by the $G$\ magnitude and test the hypothesis in each bin separately. We use bin edges of 12.5, 16.2, 17, 17.6, 18.3, and 19.3 $G$ manitude for the binning. 

\begin{figure}\centering
	\includegraphics[width=0.9\linewidth]{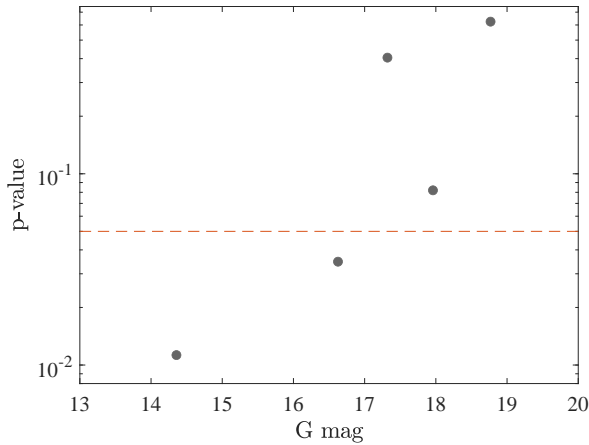}\hfill
	\caption{The $p$-value of the two-sample Kolmogorov--Smirnov test for the known binaries and Gaia sample std$(\bar{\delta}_i)$, as a function of the $G$ magnitude. The samples were binned by $G$ magnitude with bin edges of 12.5, 16.2, 17, 17.6, 18.3, and 19.3. Here, the x-axis shows the middle $G$ magnitude for each bin. The red, dashed line represents $p\text{-value}=0.05$.}
	\label{fig:pvalue}
\end{figure}

Figure \ref{fig:pvalue} shows the $p$-value of a KS-test as a function of the middle $G$ magnitude of each bin. The KS-test rejects the null hypothesis for the two brightest bins, $16.2<G<17$ and $G<16.2$, with a $p$-value of $3.4\%$ and $1.1\%$, respectively. Table \ref{tab:ks_test} shows the full results of the KS-test.
This result is expected, as the Poisson noise increases for fainter objects (see Figure \ref{fig:G_prec}). Moreover, we expect brighter asteroids to show a higher astrometric wobble amplitude, as they tend to be closer.

\begin{figure}\centering
	\includegraphics[width=1\linewidth]{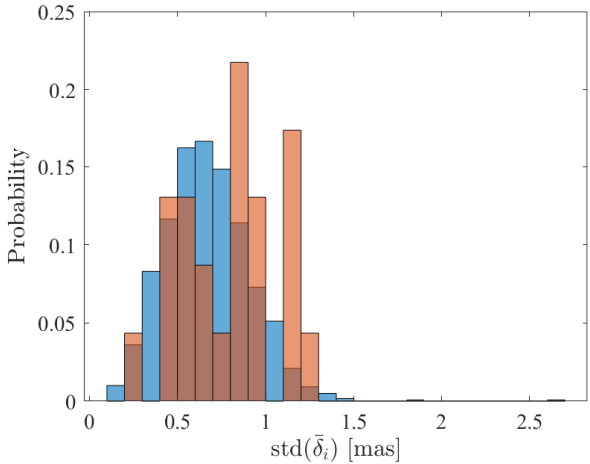}\hfill
	\caption{Histogram of the mean along-scan residuals for bright ($G<16.2$) asteroids from the known binaries' sample \citep{pravec2019_asteroid_database} (orange) and the rest of Gaia DR2 SSO objects (Blue). Both histograms normalized separately. The bin size is 0.1 mas.}
	\label{fig:amp_hist_lower_17}
\end{figure}

Figure \ref{fig:amp_hist_lower_17} shows the histograms of both samples in the brightest bin (G$<$16.2). The Gaia sample and the known binaries show a mean $\text{std}(\bar{\delta}_i)$ of 0.65 and 0.79 milliarcseconds, respectively. Therefore, we observe a slight and marginally significant excess in known binaries' residuals from the orbital fit around the Solar System barycenter. This excess in the astrometric signal suggests that binary asteroids are detectable using Gaia data.

\section{simulations}
\label{sec:simulations}

In this section, we investigate the binary asteroid detection sensitivity of the astrometric method under different levels of Gaia-like sampling and astrometric noise. 
To do so, we simulate along-scan astrometric residuals from the binary center-of-light wobble around the center-of-mass and run our algorithm to find the best fit orbital parameters.

All of the simulations in this section were conducted using the observed binary orbital parameters of (762) Pulcova, taken from \cite{main_belt_orbits_ao} and shown in Table \ref{tab:orbital_ref}.

When applying the astrometric method to actual data, we break the fitting procedure of each candidate into two steps to save computing time. In the first step, we fit the candidate residuals from the Solar System orbital fit to a circular-orbit model (i.e., set $\omega=0$ and $e=0$). Then, in the second step, we apply the entire algorithm to an elliptical-orbit model (i.e., $\omega$ and $e$ as free parameters) over candidates that show prominent FPRs (i.e., small $\alpha$) for the circular model. 

We use simulations to justify the circular simplification. First, we simulate a signal of (762) Pulcova with the Gaia DR2 SSO sampling using an elliptical-orbit model. We set the eccentricity to $e=0.3$ and add an independent Gaussian noise with a zero mean and an std of 0.4\,mas ($\mathord{\sim}$10\% of the wobble amplitude). Then, we use our pipeline to fit the elliptical- and circular-orbit, separately.

\begin{figure}\centering
	\begin{center}
		\includegraphics[width=1\linewidth]{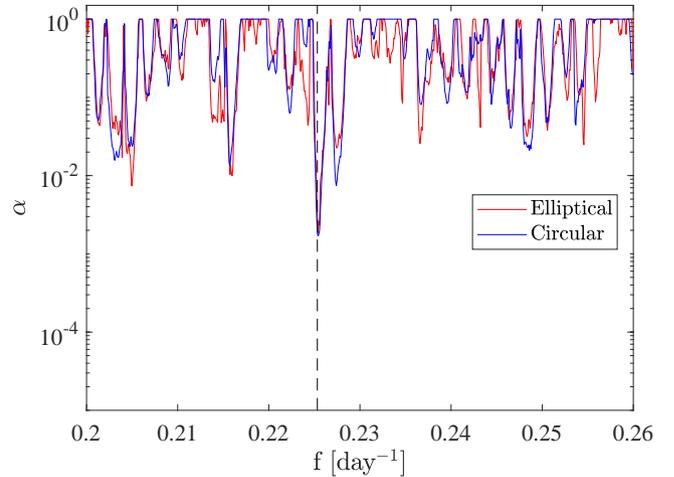}
		\hfill
		\caption{Comparison of the FPR as a function of frequency for both the elliptical- and circular-orbit fit to simulated residuals of an elliptical-orbit with high eccentricity ($e=0.3$).}  
		\label{fig:ellip_vs_circ}
	\end{center}
\end{figure}

Figure \ref{fig:ellip_vs_circ} shows the comparison between the elliptical and circular fits, where the red and blue lines are the FPR  of the elliptical and circular fits, respectively.
The circular fit restores the orbital period and shows a slight deviation in the FPR with respect to the full elliptical fit.

\begin{figure*}\centering
	\includegraphics[width=0.9\linewidth]{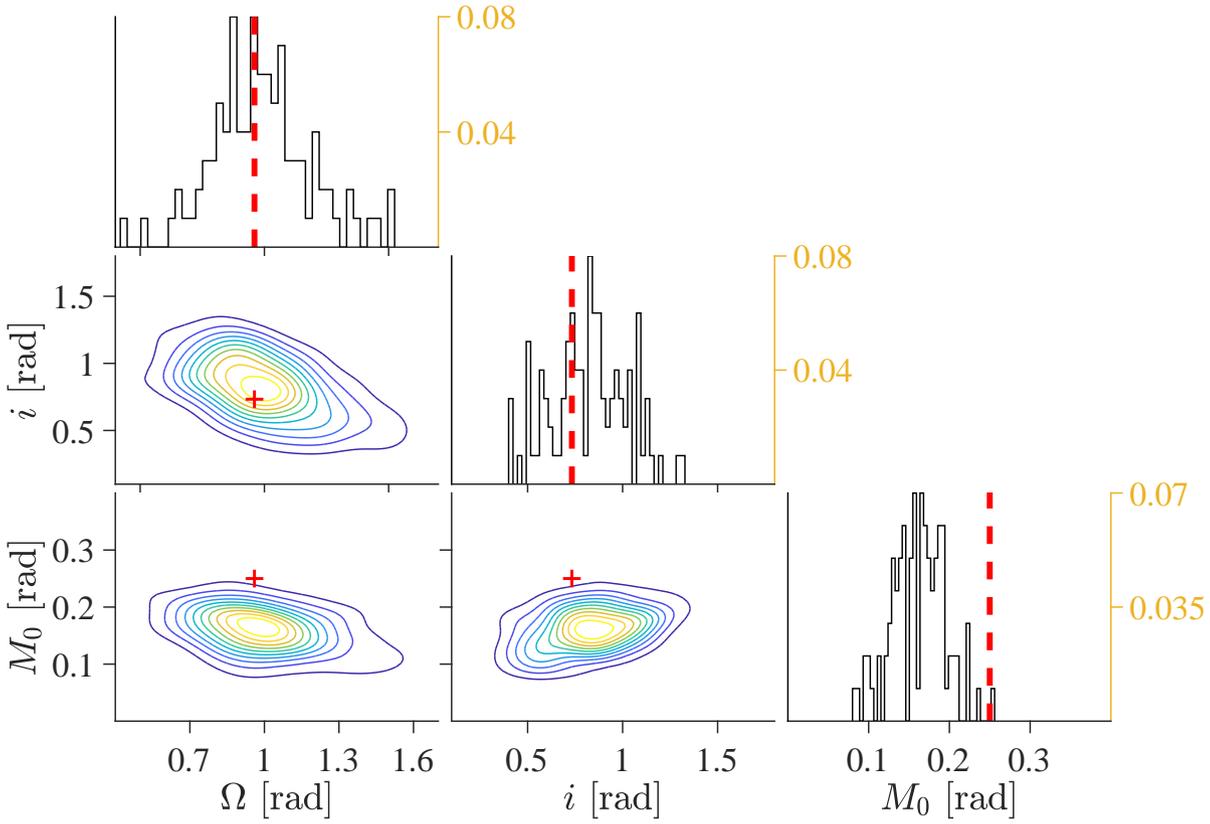}\hfill
	\caption{A corner plot with the distribution of the fitted circular-orbit model parameters over 100 simulations, where the elliptical-orbit signal is calculated for (762) Pulcova-like parameters as in Table \ref{tab:orbital_ref}, but with a high eccentricity of $e=$0.3. We add Gaussian noise with a zero mean and a standard deviation of $\sigma=0.4$\,mas  ($\sigma/\mathcal{A}\simeq 0.1$) to the simulated signal . The true parameters' values are indicated by red crosses and dashed lines.}
	\label{fig:corner_circ}
\end{figure*}

\begin{figure*}\centering
	\includegraphics[width=0.9\linewidth]{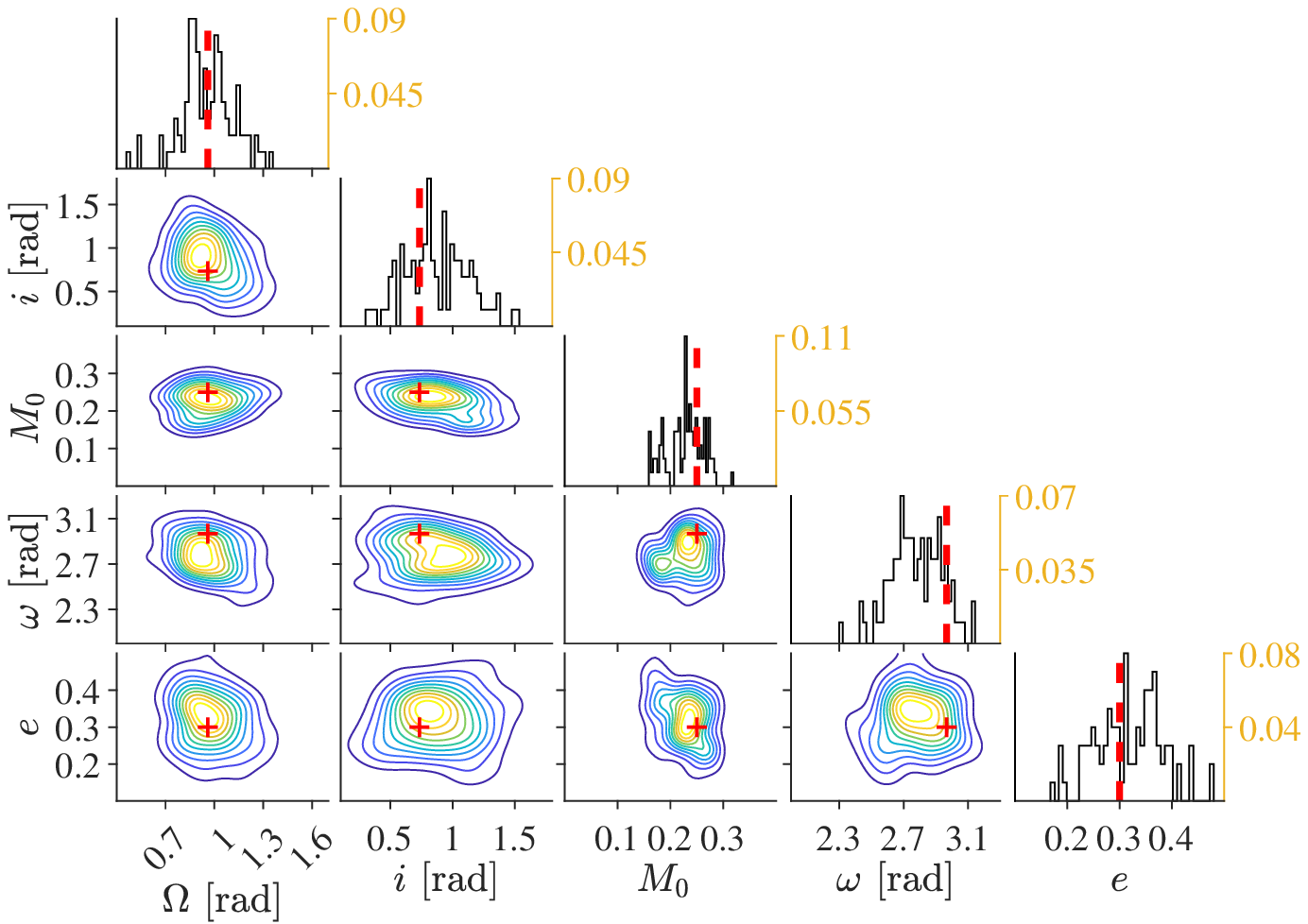}\hfill
	\caption{Same as in Figure \ref{fig:corner_circ}, but for the elliptical orbit fit.}
	\label{fig:corner_ellip}
\end{figure*}

The best fit parameter distributions from the circular- and elliptical-orbit fits are presented in Figure \ref{fig:corner_circ} and \ref{fig:corner_ellip}, respectively. 
The actual values (red crosses and dashed lines) of the longitude of the ascending node ($\Omega$) and inclination ($i$) are found within about one standard deviation from the mean values of the circular fit. However, the mean anomaly in the J2014.0 distribution is shifted in relation to the actual value. The explanation for this shift is a contribution from the argument of periapsis ($\omega$), which rotates the orbit in a relatively similar way.

\begin{figure}
	\begin{center}
		\includegraphics[width=0.9\linewidth]{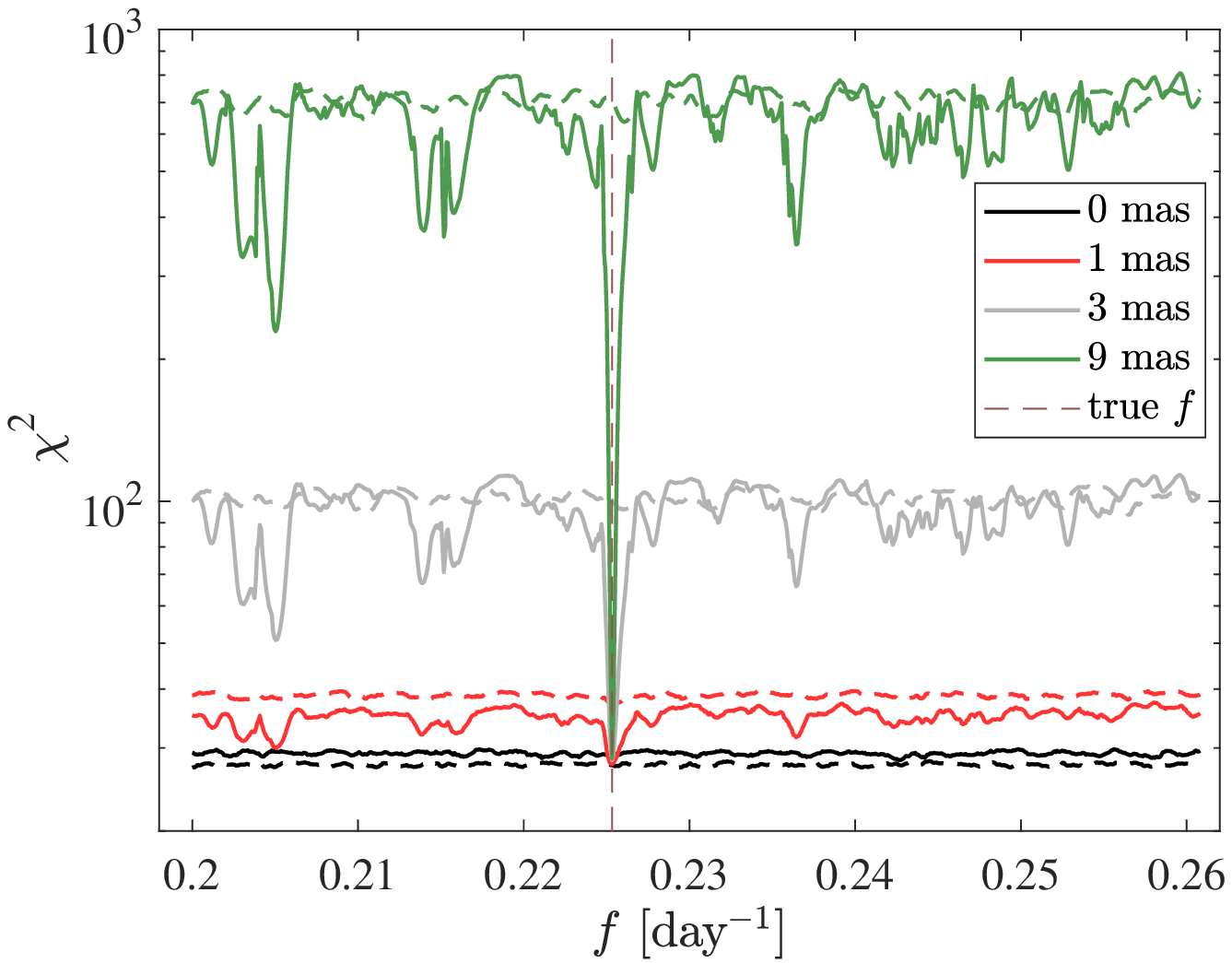}\hfill
		\caption{$\chi^2$ as a function of the frequency for a simulated (762) Pulcova with an astrometric noise of 1\,mas calculated for the Gaia DR2 sampling. The red, grey, and green solid lines are the simulated astrometric signal with an amplitude of 1, 3, and 9\,mas, respectively. The dashed lines represent the mean over 100 Bootstrap simulations for each amplitude level.} 
		\label{fig:example_chi_sq}
	\end{center}
\end{figure}

We start by presenting the results from one set of simulations. 
In these simulations, we add an independent Gaussian noise with a zero mean and a standard deviation of 1\,mas to Gaia DR2 time sampling-simulated residuals of (762) Pulcova with different wobble amplitudes. The solid and dashed lines in Figure \ref{fig:example_chi_sq} show the $\chi^2$ for the binary fit (i.e., $\chi^2(H_1)$) and the Bootstrap mean over 100 resamplings (i.e., the mean of $\chi^2(H_0)$), respectively.

We perform another set of simulations to investigate the astrometric method's sensitivity as a function of the number of data points and noise.
To ensure a realistic time sampling law, we use the actual times of the observations
of some known asteroids observed by Gaia.
When we test time spans longer than the Gaia DR2 1.8 years of observations,
we attach several epochs of observations of different asteroids into one time-series. We fit over the same frequency grid for both Gaia DR2 and eight years of Gaia-like simulations.
This simulations does not take into account the additional expected improvement in the Gaia astrometric solutions due to better modeling of the data.

\begin{figure}
	\begin{center}
		\includegraphics[width=0.9\linewidth]{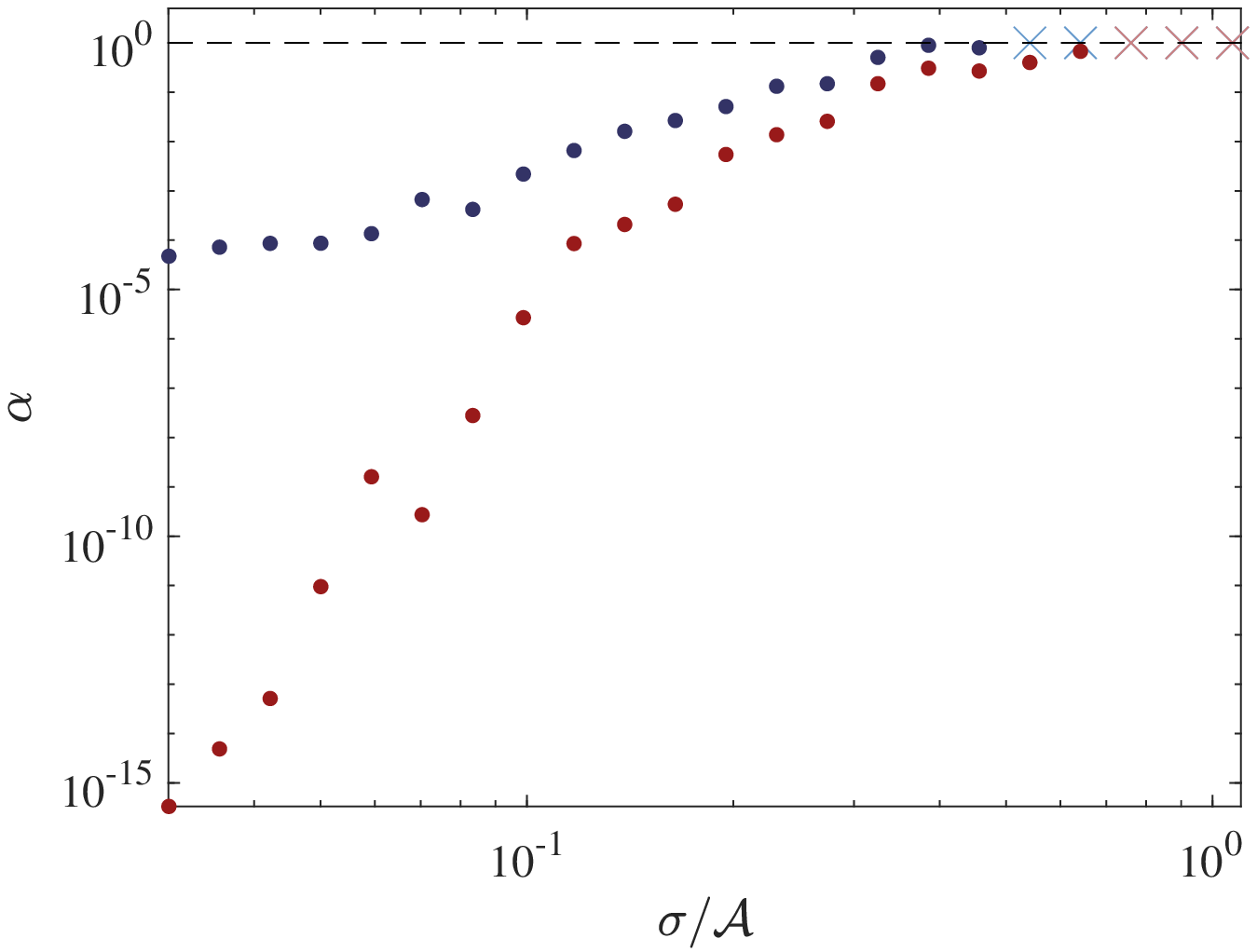}\hfill
		\caption{The best fit false-positive rate of the simulated (762) Pulcova, for both Gaia DR2 and Gaia-like time samplings, as a function of the injected astrometric noise.
		The blue (red) dots represent the mean false alarm rate for the best fit model, calculated over 100 simulations, for the Gaia DR2 (8-year Gaia-like) time sampling. Simulations in which the algorithm failed to recover the (762) Pulcova's period are marked by crosses.}
		\label{fig:Delta_chi_sigma}
	\end{center}
\end{figure}

Next, we simulate a center-of-light wobble and add a Gaussian noise with a zero mean and several standard deviation values.
Figure \ref{fig:Delta_chi_sigma} shows the FPR as a function of the astrometric noise, where the red and blue dots are calculated by Gaia DR2 and the eight-year Gaia-like sampling, respectively.
The Gaia-like sampling is generated by combining epochs from the Gaia DR2 SSO of (762) Pulcova, (90) Antiope, (267) Triza, and (670) Ottegebe.
The eight-year Gaia-like sampling shows improved sensitivity to astrometric noise by a factor of two with respect to the Gaia DR2 SSO sampling, as shown in Figure \ref{fig:Delta_chi_sigma}. Therefore, applying the astrometric method to the next Gaia SSO data release has the potential to improve the detection sensitivity.

\begin{table*}
\centering
\caption{Summary of the orbital parameters of (762) Pulcova, taken from \protect \cite{main_belt_orbits_ao}.}
\label{tab:orbital_ref}
\begin{tabular}{|l|c|}
\hline\hline
  &    (762) Pulcova  \\
  \hline
  Period (days) &4.438$\pm$0.001\\
  Semi-major axis (km) & 703 $\pm$ 14 \\
  Eccentricity &  0.03$\pm$ 0.01 \\
  $r_1/r_2$ &0.134$\pm$ 0.049\\
  Inclination in J2000 (deg) & 132$\pm$2\\
  Argument of periapsis (deg) &170$\pm$20\\
  Time of periapsis (JD) &  2453813.5$\pm$ 0.012 \\
  Ascending node (deg) & 235$\pm$2 \\
\hline
\hline
System mass (kg) &  1.40$\pm 0.1\times
 10^{18}$\\
 \hline
 \hline
\end{tabular}

\end{table*}

\section{Gaia DR2 results}
\label{sec:results}
In this section, we apply our algorithm to data from the Gaia DR2 SSO catalog.
We focus our search on selected candidates that satisfy the following criteria:
\begin{enumerate}
    \item rstd$(\bar{\delta})>1.5$\,mas, where rstd is the robust standard deviation\footnote{\label{foot:rstd}We define the robust standard deviation (rstd) as $\frac{1.4826}{2}\left(Q_3-Q_1\right)$, where $Q_1,Q_3$ are the first and third quartiles, respectfully.}.
    \item Number of epochs in Gaia DR2 $>12$.
\end{enumerate}

A total of 314 asteroids satisfied those criteria, out of which we choose the 20 asteroids with the highest rstd$(\bar{\delta})$/mean$(\sigma)$ ratio (shown in Table \ref{tab:result_gaiadr2}), where two are known binary systems from \cite{johnston2018binary}, (5899) Jedicke and (2131) Mayall, and one is a triple system: (93) Minerva, Aegis, and Gorgoneion. 

In addition, we apply the astrometric method on the (4337) Arecibo binary system \citep{arecibo_sat_dectection_gault2022new}, which shows an astrometric signal in \cite{tanga2022data_GAIA_DR3_SSO} but did not satisfy the criteria listed above. The results for (4337) Arecibo are presented in \S\ref{subsec:arecibo}.

For each candidate, we fit binary orbital parameters for frequencies ranging from 1/350\,\,$\text{hr}^{-1}$ to 1/10\,\,$\text{hr}^{-1}$, with steps of 1/416\,\,$\text{hr}^{-1}$. 
Next, we run Bootstrap simulations with 500 resamplings in order to calculate the FPR, as described in \S \ref{sec:model_inversion}.

\begin{figure}\centering
	\begin{center}
		\includegraphics[width=0.9\linewidth]{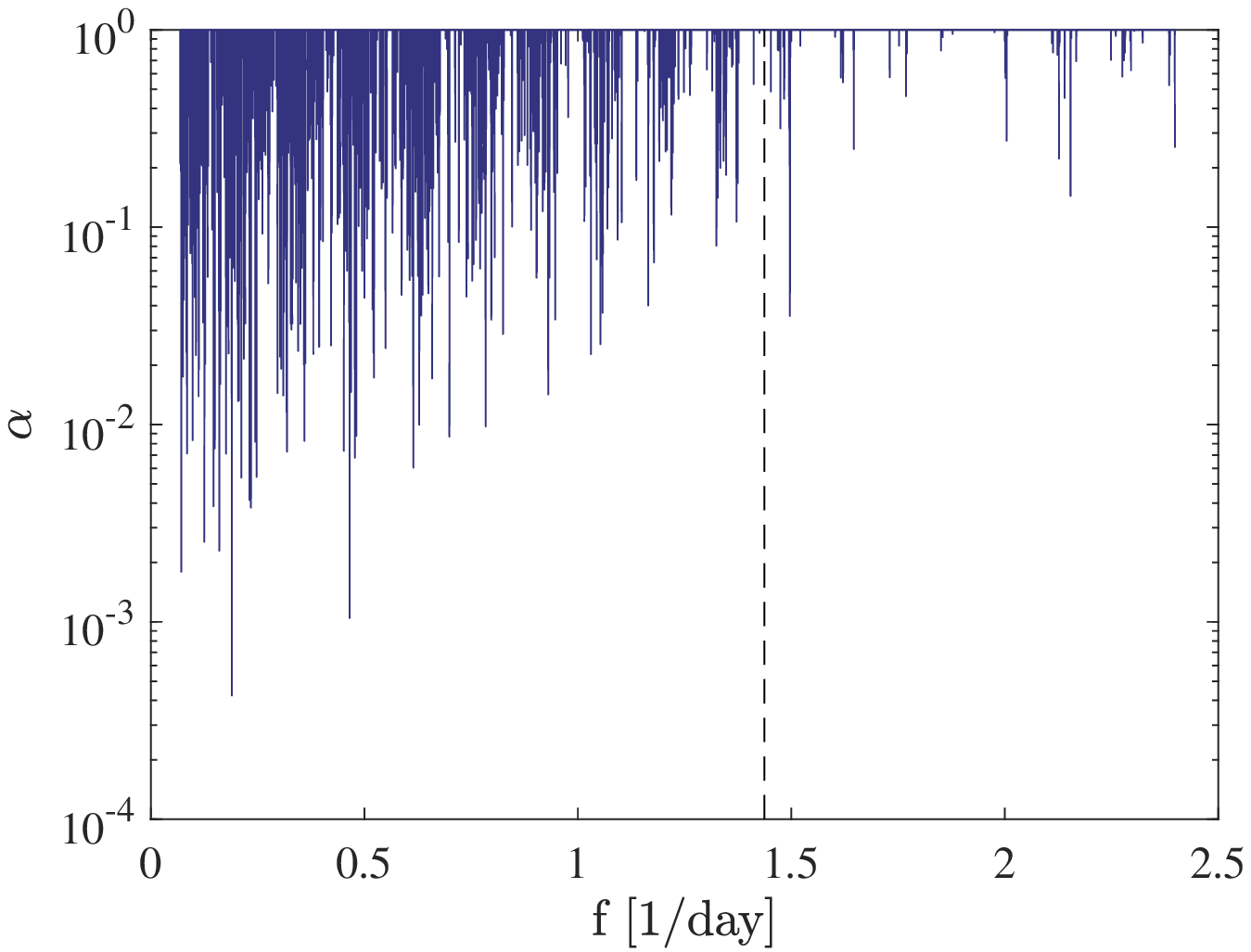}\hfill
		\caption{The false-positive rate (per frequency) of the best fit for Gaia DR2 data of the known binary asteroid (5899) Jedicke as a function of frequency. To determine the false-positive rate, we calculate the null hypothesis distribution using 500 Bootstrap resamplings (see \S\ref{sec:model_inversion}). The vertical dashed line marks the observed orbital period from \protect \cite{johnston2018binary}.} 
		\label{fig:fpr_5899}
	\end{center}
\end{figure}

\begin{figure}\centering
	\begin{center}
		\includegraphics[width=1.07\linewidth]{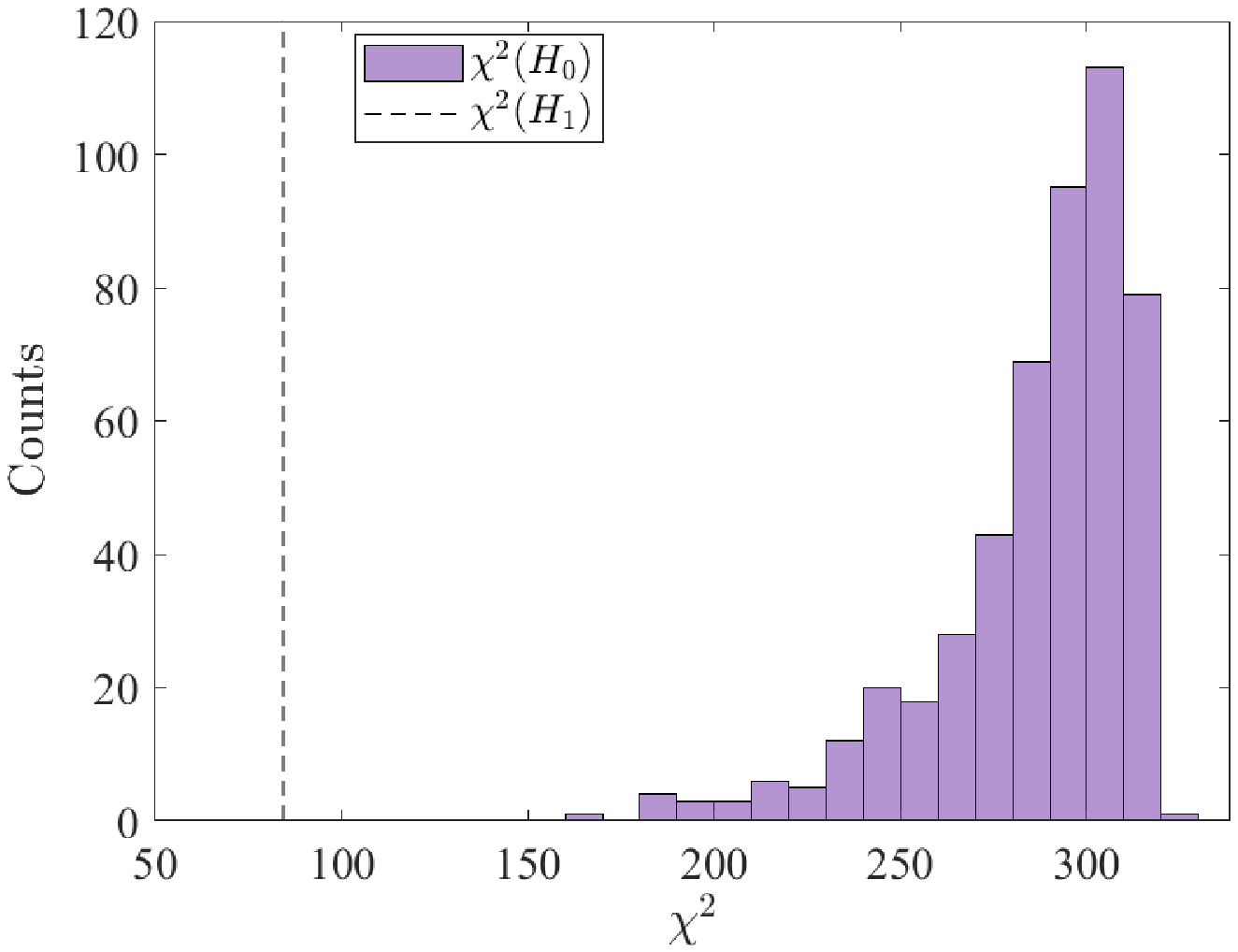}\hfill
		\caption{The null hypothesis $\chi^2(H_0)$ distribution for (5899) Jedicke for the best fit frequency for the Gaia DR2 SSO catalog (shown in Table \ref{tab:result_gaiadr2}). The null hypothesis distribution were calculated by 500 Bootstrap resamplings (see \S\ref{sec:model_inversion}). The black dashed line shows $\chi^2(H_1) = 83.03$, the best fit for (5899) Jedicke in Gaia DR2.} 
		\label{fig:best_fit_h0_dist}
	\end{center}
\end{figure}

The results show no significant binarity in the selected 20 asteroids. We present an example of the fit results for (5899) Jedicke, a known binary asteroid in the main belt with 14 epochs in Gaia DR2.
The FPR as a function of the orbital frequency for a binary model fit is shown in Figure \ref{fig:fpr_5899}, and the best fit $\chi^2(H_1)$ together with the $\chi^2(H_0)$ distribution for the best fit frequency are shown in Figure  \ref{fig:best_fit_h0_dist}.

\begin{figure}
\centering
	\begin{center}
		\includegraphics[width=0.9\linewidth]{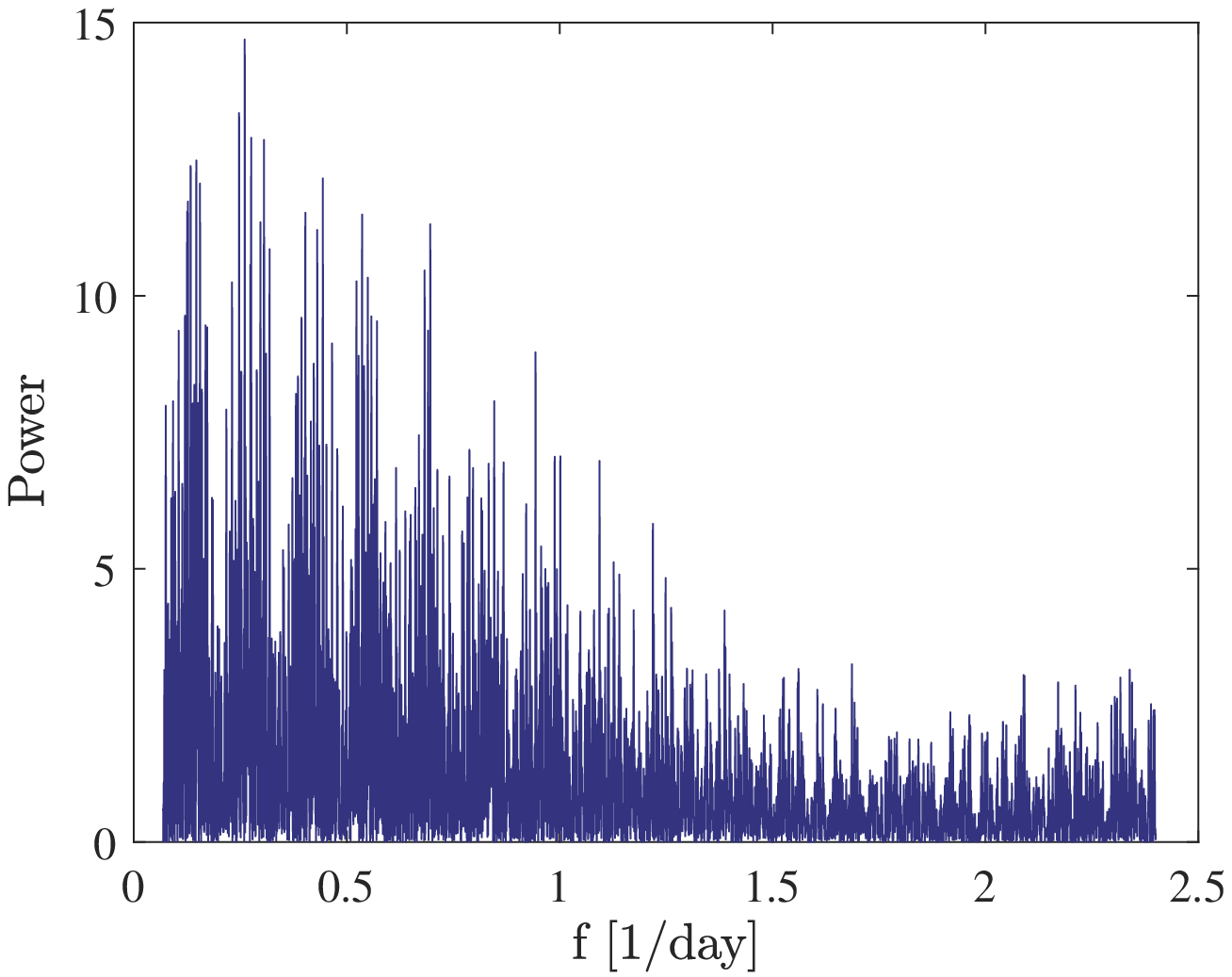}\hfill
		\caption{The window function spectrum for the known binary asteroid (5899) Jedicke observations in Gaia DR2.  } 
		\label{fig:window_26367}
	\end{center}
\end{figure}

 Figure \ref{fig:window_26367} shows the window function for the observations of (5899) Jedicke in the Gaia DR2 SSO, calculated by 
\begin{align}
W(f) = \left|\sum_{j} \, e^{-2\uppi i f t_j}\right|^2,
\end{align}
where the $j$-index represents observation and $t$ is the observation time.
The window function power spectrum shows complicated features. The Whittaker--Kotel'nikov--Shannon (WKS) sampling theorem describes a discrete time series as a convolution of the true continuous signal with the window function. In this paper, we invert the problem to fit a continuous binary model to the discrete time-sampling of Gaia DR2 SSO observations. Unfortunately, the convolution with a complicated and non-regular window function generates severe aliasing, which generates complex frequency correlations. Those correlated frequencies limit the ability to fit a periodic signal to the data set. Using more extended baseline observations and a complete dataset, one that does not clip outliers, may improve the aliasing and increase the probability of detecting new binary asteroids in future Gaia data releases.

\subsection{(4337) Arecibo}
\label{subsec:arecibo}
The satellite of (4337) Arecibo was first detected in May 2021 by stellar occultation \citep{arecibo_sat_dectection_gault2022new}. \cite{tanga2022data_GAIA_DR3_SSO} demonstrate the astrometric signal of (4337) Arecibo in Gaia DR3 SSO catalog by plotting the along-scan residuals from the single object Solar Barycenter orbital fit for 13 transits (epochs), which spanned over 2.3 days. \cite{tanga2022data_GAIA_DR3_SSO} report an orbital period of $32.85\pm0.38$ hours, based on the light curve from "Behrend, R. et al., private communication", which did not publish at the time of writing this paper. The observed physical parameters are listed in Table \ref{tab:Arecibo}.

We apply the astrometric method to the Gaia DR2 SSO observations of the binary system (4337) Arecibo. We fit for a circular binary orbit, for frequencies ranging from 1/350\,hr$^{-1}$ to 1/10\,hr$^{-1}$, with steps of 1/2080\,hr$^{-1}$, and run Bootstrap simulations with 500 resamplings.

\begin{figure}\centering
	\begin{center}
		\includegraphics[width=0.9\linewidth]{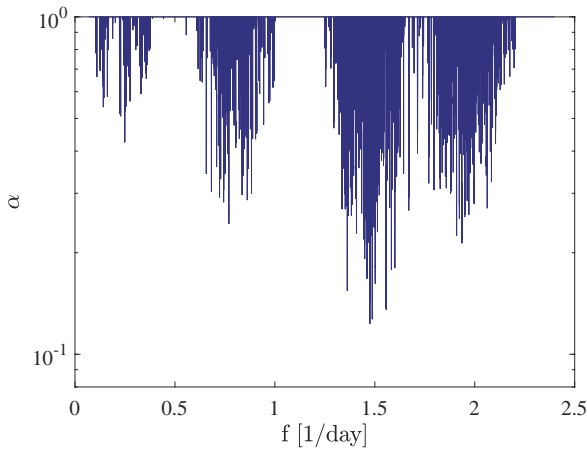}\hfill
		\caption{The false-positive rate (per frequency) of the best fit for Gaia DR2 data of the known binary asteroid (4337) Arecibo as a function of frequency. To determine the false-positive rate, we calculate the null hypothesis distribution using 500 Bootstrap resamplings (see \S\ref{sec:model_inversion}). } 
		\label{fig:fpr_arecibo}
	\end{center}
\end{figure}

\begin{figure}
\centering
	\begin{center}
		\includegraphics[width=0.9\linewidth]{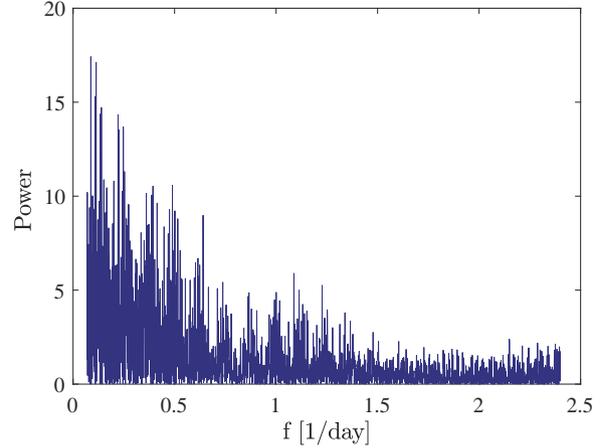}\hfill
		\caption{The window function spectrum for the known binary asteroid (4337) Arecibo observations in Gaia DR2.  } 
		\label{fig:window_arecibo}
	\end{center}
\end{figure}

Figures \ref{fig:fpr_arecibo} and \ref{fig:window_arecibo} show the best fit FPR as a function of frequency and the window function for (4337) Arecibo's Gaia DR2 observations. We report the best period of $P = 16.26$\,hours, which is half the period reported in \cite{tanga2022data_GAIA_DR3_SSO}. However, the FPR for the best fit is $\mathord{\sim}$0.11, and therefore the fitted astrometric model is insignificant according to our estimation of the FPR. 

\begin{figure*}
	\includegraphics[width=0.9\linewidth]{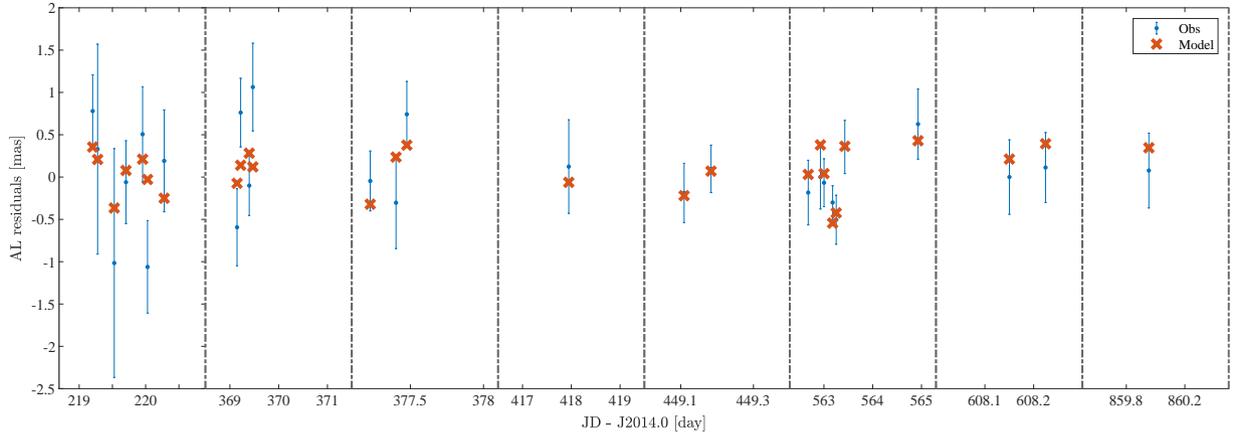}
	\caption{The observed (4337) Arecibo Gaia DR2 along-scan (blue dot) residuals plotted together with the best fit model (orange cross). The error bar of the Gaia DR2 along-scan residuals was calculated by Equation \ref{eq:sigma_i}. The sixth section from the left (JD-J2014.0 of 562.5 to 565 days) contains the 2.3 days time range presented in \protect \cite{tanga2022data_GAIA_DR3_SSO}. There are only seven Gaia DR2 SSO epochs in this time range, while there are 13 epochs in Gaia DR3 SSO.}
	\label{fig:arecibo_oc}
\end{figure*}

The observed Gaia DR2 along-scan residuals and the calculated best fit are plotted in Figure \ref{fig:arecibo_oc}. Due to the high FPR and the incompatible residual model, we do not reject the null hypothesis, i.e., we do not observe an astrometric signal in the (4337) Arecibo observations in Gaia DR2. The 2.3\,days of observations presented in \cite{tanga2022data_GAIA_DR3_SSO}, which contains 13 epochs in Gaia DR3, contains only seven epochs in Gaia DR2 (see Figure \ref{fig:arecibo_oc}).

\begin{table*}
\centering
\caption[caption without footnote]{Summary of the astrometric method test and physical parameters for (4337) Arecibo.}
\label{tab:Arecibo}

\begin{tabular}{|c|c|c|c|c|c|c|c|c|c|c|c|c|}

\hline\hline
\noalign{\vskip 0.5mm}    
    \parbox{1cm}{ MPC index}  &  \parbox{1.1cm} {\# Gaia \\epochs } &    \parbox{1cm}{mag G}   & \parbox{1cm}{ rstd($\bar{\delta}$) \\\centering(mas)}   & \parbox{1cm}{ Mean($\sigma$)\\ (mas)}  & \parbox{1cm}{ $D_1$\\ (km)} & \parbox{1cm}{ $D_2$\\ (km)}  &  \parbox{1cm}{Ref. $P$\\ (hour)}  & \parbox{1cm}{Exp. $\mathcal{A}$\\ (mas)}  &$P$ (hour)  &     $\alpha$ (FPR) \\  
\hline
\noalign{\vskip 0.5mm}
      4337     &     27    &   17.8    &   0.51  &   0.48    &   24.4$\pm$0.6  &    13.0$\pm$1.5 &32.85$\pm$0.38      &      2.01      &    16.26  &   0.11   \\

 \hline
 \hline
\end{tabular}

     {\raggedright The components diameters ($D_1$ and $D_2$) were taken from \cite{arecibo_sat_dectection_gault2022new}. The reference period ("Ref. $P$") are mentioned in \cite{tanga2022data_GAIA_DR3_SSO}. Both the diameter ratio and the reference catalog were used for the expected amplitude calculation ("Exp. $\mathcal{A}$") in Equation \ref{eq:col_q}, where we use the mean Gaia-to-target distance to convert into angular distance and binary separation of 49.9\,km \citep{tanga2022data_GAIA_DR3_SSO}. $P$ is the best fit period, and $\alpha$ is the related minimal false-positive rate for the best fit. \par}
\end{table*}

\section{caveats}
\label{sec:caveats}

In this section, we describe observational and physical caveats that may limit the performance of the astrometric method. 

\subsection{Removal of outliers}
The main observational caveat of the Gaia DR2 SSO catalog is the removal of outliers from the published catalog. Unfortunately, this step excludes the most informative observations of the astrometric method. Therefore, in order for this method to be used successfully, the full data set, without outliers removal, is needed.

\subsection{Orbital fitting}
Currently, we fit the asteroid binary model post Solar System orbital fit. The center-of-light astrometric wobble is a periodic signal with a zero mean. With a finite number of epochs, the center-of-light wobble will add a systematic error to the Solar System orbital fit. The amplitude of this systematic error is of the order of $\mathcal{A}/\sqrt{N_e}$, where $N_e$ is the number of epochs. For $N_e\mathord{\approx} 25$, which is typical for the Gaia DR2 SSO catalog, this translates into $\mathord{\sim}0.2$ of the astrometric wobble amplitude. Although this effect is expected to decrease in future data releases, \text{it may be }worth fitting the Solar System orbit and the center-of-light wobble simultaneously.

\subsection{Asteroid rotation}
Variations in the light curve of a single asteroid can contaminate the astrometric wobble signal. We consider two types of effects: (i) A binary system in which the center-of-light changes due to variability in one or two of the components' flux (e.g., due to the rotation of a highly elongated component). (ii) A single asteroid whose center-of-light changes during rotation (e.g., due to a non-homogeneous albedo).

A previous study of the shape distribution of asteroids that appear in the Gaia DR2 SSO catalog \citep{aspect_ratio_mommert2018main} shows that large asteroids ($D\mathord{>}50$km) in the main belt are slightly elongated with a mean aspect ratio of $b/a\,\mathord{\approx}\,0.86$. The smaller asteroids in the sample of \cite{aspect_ratio_mommert2018main} have a mean aspect ratio of $b/a\,\mathord{\approx}\,0.79$, and include a non-negligible number of asteroids with $b/a\,\mathord{\approx}\,0.5$. 

If the variable asteroid is in a binary system, this rotational variation can bias the binary orbital parameters determined by the astrometric method. For example, in the case of an elongated secondary, the center-of-light position will change periodically; this is because the secondary flux will vary while the primary flux will remain constant, or vice versa. This effect may contaminate the astrometric wobble due to orbital motion in the binary system with an additional periodic signal. The amplitude of this additional periodic signal is approximately linear with the secondary-to-primary flux ratio. Therefore, we expect an amplitude of $\mathord{\sim} \left(b/a\right)\mathcal{A}$.
This doubly (or triple, if both components are variables and a-synchronous) periodic signal will be observed by the astrometric method and may be distinguished using photometric observations that reveals the individual rotation periods.

A periodic astrometric shift of the center-of-light may be generated in the case of a single asteroid with an asymmetric shape or albedo surface variation. The astrometric method may falsely detect this shift as a binary asteroid.
For example, in an extreme and non-realistic case of an asteroid whose fluxes all originate from a small patch on the equator, the astrometric shift amplitude will be of the size of the equatorial asteroid radius. The expected amplitude in this extreme case, for a main belt asteroid whose diameter is 10\,km at a distance of 2\,au is $\mathord{\sim}$3\,mas.
Fortunately, the amplitude of this effect is orders of magnitude lower for asteroids with realistic albedo variations and can be assessed by monitoring their lightcurve.

\subsection{Precession}
The precession of a binary asteroid orbital plane adds a periodic signal to the position of the center-of-light, which requires additional parameters to be modeled.

A first-order approximation of the nodal and apsidal precession angular rate is given by (e.g., \citealt{greenberg1981apsidal_precession})
\begin{align}
\omega_p \simeq \frac{3\pi}{P} \frac{R^2}{a^2} J_2,
\label{eq:precession}
\end{align}
where $P$ is the orbital period, $R$ is the primary radius, $a$ is the orbital semi-major axis, and $J_2$ the gravitational second harmonic coefficient of the primary component. Note that the determination of $J_2$ requires a model of the asteroid's shape. In the case of an ellipsoid with semi-major axis $a$, semi-intermediate axis $b$, semi-minor axis c (i.e., $a>b>c$) and constant density, the $J_2$ coefficient can be approximated by:
\begin{align}
J_2\approx\frac{1}{5} \frac{a^2+b^2-2c^2}{a^2},
\end{align}
which translate into $J_2\approx 0.05$ for the mean asteroids with $b/a=0.79$ in the sample of \cite{aspect_ratio_mommert2018main}, where we take $c \approx b$.
For example, by inserting the orbital parameters and the shape model of the main belt binary (22) Kalliope, where $a/b=1.32$ and $b/c=1.2$ \citep{kalliope_modelmarchis2003three_}, we obtain a $J_2\approx 0.15$ and expected precession period of $\mathord{\sim} 7.5$ years.
However, a previous adaptive optics campaign (see \citealt{main_belt_orbits_ao}), which spanned over $\mathord{\sim} 5$ years of observation, did not detect any nodal precession for 22 Kalliope and 762 Pulcova.
Nevertheless, the precession needs to be added to our model when applying the astrometric method to future data releases of Gaia, which contains observations over a longer time baseline.

\section{conclusion}
\label{sec:conclusion}

We present a method for detecting binary asteroids based on the motion of the center-of-light of a binary (or primary) asteroid around the center-of-mass. We derive a forward model for the center-of-light wobble around the center-of-mass as seen by the observer and describe a procedure to invert the problem and fit the binary orbital parameters.

In \S\ref{sec:simulations}, we investigate the performance of the astrometric method. Our result suggests that binary asteroid detection may be feasible in future data releases even under the current performance of Gaia DR2, without considering the improvement in the Gaia data reduction. However, this will likely require all of the measurements, including the outliers that were not used (and not published) in the orbital fit.

We present the known binary asteroids' population that appears in the Gaia DR2 Solar System catalog in \S \ref{sec:data_collection}. 
The data-reduction procedure rejects outliers from the orbital Solar System fit, although the measurements are consistent between sub-transits (see \S\ref{sec:data_collection}). Unfortunately, this rejection procedure excludes the most informative observations from the input data used in our pipeline. Therefore, it is important to publish these outliers' data points in order to search for binary asteroids using the astrometric method.
Interestingly, using the KS-test, we found that the known binary asteroids show a slight, marginally significant excess in the Solar System orbital fit residuals relative to the rest of the asteroids.

Applying our astrometric method to 20 selected asteroids did not lead to a significant detection of a binary asteroid. Three of the selected asteroids were known as multiple asteroid systems.

In addition, we apply the astrometric method to the (4337) Arecibo observations in Gaia DR2. About half (6 out of 13) of (4337) Arecibo's epochs on the Gaia DR3 SSO subset which were presented in \cite{tanga2022data_GAIA_DR3_SSO}, were not published in Gaia DR2 SSO. Gaia DR2 SSO measurements of (4337) Arecibo reveal a marginally significant peak period of half the reported period \citep{tanga2022data_GAIA_DR3_SSO}.

Future data releases that contain all of the Gaia observations and a more extended time baseline will improve the detection sensitivity of the astrometric method.
\section*{Acknowledgements}

E.O.O. is grateful for the support of grants from the Willner Family Leadership Institute, André Deloro Institute, Paul and Tina Gardner,
The Norman E Alexander Family M Foundation ULTRASAT Data Center Fund, Israel Science Foundation, Israeli Ministry of Science, Minerva, BSF, BSF-transformative,
NSF-BSF, Israel Council for Higher Education (VATAT), Sagol Weizmann-MIT, Yeda-Sela, and Weizmann-UK.
\section*{Data availability}
The data underlying this article are available in the Gaia archive, at \url{https://cdn.gea.esac.esa.int/Gaia/gdr2/}.

\bibliographystyle{apalike}

\begin{appendices}
\section*{ GAIA's along-scan projection}
\label{subsec:gaia_al}
We apply the astrometric method to detect binaries on data from Gaia DR2 (see \S\ref{sec:data_collection}).
We project the 2D center-of-light vector to the Gaia along-scan axis using the Gaia position angle.

The Gaia position angle ($PA$) describes the spacecraft scanning direction in the equatorial reference frame.
The position angle is defined such that $PA=0 $ represents rotation toward the equatorial north pole that increase toward the east.
To extract the along-scan component from the 2D center-of-light position $\vec{x}_{col}$, with $\hat{x}_z$ pointing to the equatorial north pole, we apply the following projection
\begin{align*}
    x_{al} = \vec{x}_{col}\cdot\left[\cos\left(PA\right)\hat{n}_N+\sin\left(PA\right)\hat{n}_E\right],
\end{align*}
where $\hat{n}_N$ and $\hat{n}_E$ are defined in Equations \ref{eq:hat_nN} and \ref{eq:hat_nW}, respectively.

\section*{}

\begin{table*}
\centering
\caption{Summary of the two-sample Kolmogorov--Smirnov test for the known binaries \citep{pravec2019_asteroid_database} and Gaia DR2 sample std$(\bar{\delta}_i)$, for binned $G$ magnitude.}
\label{tab:ks_test}
\begin{tabular}{|l|c|c|c|}
\hline\hline
  Edges [G mag]& $p$-value \% & Number of Gaia objects &Number of Parvec objects\\
  \hline
  12.5-16.2 & 1.1 & 1193& 23 \\
  16.2-17.0 & 3.4 & 1189& 24 \\
  17.0-17.6 & 40.4 & 2183& 23 \\
  17.6-18.3 & 8.1 & 4086& 24 \\
  18.3-19.2 & 62.4 & 4815& 23 \\
 \hline
 \hline
\end{tabular}

\end{table*}

\begin{table*}
\centering
\caption[caption without footnote]{Summary of the astrometric method test for selected candidates from Gaia DR2.}
\label{tab:result_gaiadr2}

\begin{tabular}{|c|c|c|c|c|c|c|c|c|c|c|}
\hline\hline
\noalign{\vskip 0.5mm}    
    \parbox{1cm}{ MPC index}  &  \parbox{1.1cm} {\# Gaia \\epochs } &    \parbox{1cm}{mag G}   & \parbox{1cm}{ rstd($\bar{\delta}$) \\\centering(mas)}   & \parbox{1cm}{ Mean($\sigma$)\\ (mas)} &   \parbox{1cm}{Ref. $P$\\ (day)}  & \parbox{1cm}{Exp. $\mathcal{A}$\\ (mas)}  &$P$ (day)  &     $\alpha$ (FPR) \\  
\hline
\noalign{\vskip 0.5mm}
        93    &    26   &      13.5  &      1.19   &     0.16   &  2.408/1.115    &  0.23/0.11  &    2.815  &       0.002   \\
       118    &    22   &      12.7  &      1.15   &     0.16   &         -       &      -      &    0.949  &       0.015   \\
       250    &    16   &      12.3  &      1.23   &     0.13   &         -       &      -      &    0.426  &       0.031   \\
       346    &    14   &      12.4  &      1.38   &     0.16   &         -       &      -      &    0.743  &        0.03   \\
       386    &    15   &      12.5  &      1.37   &     0.23   &         -       &      -      &     0.56  &      0.0088   \\
       554    &    17   &      13.8  &      1.14   &     0.17   &         -       &      -      &    8.766  &       0.016   \\
       690    &    16   &      12.9  &      1.48   &     0.18   &         -       &      -      &    0.466  &      0.0072   \\
       893    &    13   &      14.7  &      1.25   &     0.19   &         -       &      -      &    0.537  &       0.023   \\
      1471    &    19   &      15.5  &      1.27   &     0.27   &         -       &      -      &    0.566  &       0.064   \\
      2131    &    14   &      15.7  &      1.25   &     0.33   &     0.978       &     0.85    &    1.289  &       0.016   \\
      2470    &    17   &        17  &      1.27   &      0.4   &         -       &      -      &    0.621  &      0.0092   \\
      5899    &    14   &      16.7  &      2.16   &     0.69   &     0.696       &     0.21    &    5.275  &      0.0006   \\
      6199    &    14   &      17.2  &      2.11   &     0.43   &         -       &      -      &     5.96  &       0.015   \\
      6315    &    14   &      18.2  &      3.33   &     0.94   &         -       &      -      &    3.673  &        0.16   \\
      7033    &    13   &      17.7  &      2.43   &     0.73   &         -       &      -      &   11.303  &       0.037   \\
      7825    &    13   &      18.4  &      2.75   &     0.75   &         -       &      -      &    2.285  &      0.0077   \\
      9356    &    15   &      17.9  &      1.45   &     0.44   &         -       &      -      &    6.869  &       0.011   \\
     10569    &    22   &      18.8  &      2.65   &     0.68   &         -       &      -      &   11.498  &     0.00094   \\
     11342    &    17   &      18.2  &      3.66   &     1.07   &         -       &      -      &    2.057  &       0.035   \\
     36731    &    13   &      18.1  &      2.59   &     0.67   &         -       &      -      &    5.498  &      0.0049   \\

 \hline
 \hline
\end{tabular}

     {\raggedright We use \cite{johnston2018binary} database for the reference period ("Ref. P") and the diameter ratio for the expected amplitude calculation  ("Exp. $\mathcal{A}$") in Equation \ref{eq:col_q}, where we use the mean Gaia-to-target distance to convert into angular distance. $P$ is the best fit period, $\alpha$ is the related minimal false-positive rate for the best fit.  For (93) Minerva, we present the orbital period and expected amplitude for both satellites, Aegis and Gorgoneion, respectively. \par}
\end{table*}

\end{appendices}

\label{lastpage}
\end{document}